\documentclass[
twocolumn,
twoside,
letterpaper
]{IEEEtran}

\ifCLASSINFOpdf
  \usepackage[pdftex]{graphicx}
  \DeclareGraphicsExtensions{.pdf,.jpeg,.png}
\else
\fi
\begin{document}
%
\title{Progress in the development of 
       TES microcalorimeter detectors suitable for neutrino mass measurement}

\author{A.~Giachero,
        B.~Alpert, 
        D.T.~Becker, 
        D.A.~Bennett, 
        M.~Borghesi, 
        M.~De Gerone, 
        M.~Faverzani, 
        M.~Fedkevych, 
        E.~Ferri, 
        G.~Gallucci, 
        J.D.~Gard, 
        F.~Gatti, 
        G.C.~Hilton, 
        J.A.B.~Mates, 
        A.~Nucciotti, 
        G.~Pessina, 
        A.~Puiu, 
        C.D.~Reintsema, 
        D.R.~Schmidt, 
        D.S.~Swetz, 
        J.N.~Ullom, 
        L.R.~Vale
	   
\thanks{A.~Giachero, 
        M.~Borghesi, 
        M.~Faverzani, 
        E.~Ferri and 
        A.~Nucciotti 
        are with Dipartimento di Fisica, Universit\`{a} di Milano-Bicocca and INFN - Sezione di Milano Bicocca, Milano, I-20126, Italy (e-mail: andrea.giachero@mib.infn.it).
        }
\thanks{B.~Alpert, 
        D.T.~Becker, 
        D.A.~Bennett, 
        J.D.~Gard, 
        J.A.B.~Mates,
        C.D.~Reintsema, 
        D.R.~Schmidt, 
        D.S.~Swetz, 
        J.N.~Ullom and 
        L.R.~Vale
        are with National Institute of Standards and Technology, Boulder, CO 80305, USA.
       }
\thanks{M.~De Gerone,
        M.~Fedkevych and 
        G.~Gallucci 
        are with INFN - Sezione di Genova, Genova I-16146 - Italy}
\thanks{F.~Gatti,
        is with Dipartimento di Fisica, Universit\`{a} di Genova and INFN - Sezione di Genova, Genova I-16146 - Italy}
\thanks{G.~Pessina is with INFN - Sezione di Milano Bicocca, Milano, I-20126, Italy}
\thanks{A.~Puiu is with Gran Sasso Science Institute (GSSI), I-67100 L’Aquila, Italy}

}


\maketitle

\begin{abstract}

The HOLMES experiment will perform a precise calorimetric measurement of the end point of the Electron Capture (EC) decay spectrum of $^{163}$Ho in order to extract information on neutrino mass with a sensitivity below 2\,eV. In its final configuration, HOLMES will deploy 1000 detectors of low temperature microcalorimeters with implanted $^{163}$Ho nuclei. The baseline sensors for HOLMES are Mo/Cu TESs (Transition Edge Sensors) on SiN$_x$ membrane with gold absorbers. Considering the large number of pixels and an event rate of about 300 Hz/pixel, a large multiplexing factor and a large bandwidth are needed. To fulfill this requirement, HOLMES will exploit recent advances on microwave multiplexing. In this contribution we present the status of the activities in development, the performances of the developed microwave-multiplexed readout system, and the results obtained with the detectors specifically designed for HOLMES in terms of noise, time and energy resolutions.
\end{abstract}


%
\IEEEpeerreviewmaketitle

\section{Introduction}
The experiments on the neutrino flavor oscillation phenomenon provide indisputably evidence of the non-vanishing nature of the neutrino mass~\cite{Esteban_2020}. Measuring the neutrino mass is one of the most urgent issues in particle physics. Several approaches are taken in order to assess this parameters, such as cosmological surveys~\cite{Lattanzi_2017}, searches for neutrinoless double beta decay~\cite{Dolinski_2019} and the studies of the end point of beta or electron capture (EC) decays~\cite{Drexlin_2013}.

An interesting approach suitable for a direct determination of the neutrino mass, is the calorimetric measurement of the energy released in the EC decay of $^{163}$Ho, as proposed by De Rujula and Lusignoli in 1982~\cite{DeRujula_1982}. The proximity of the $Q$-value ($Q_{\mbox{\tiny EC}}=2.883$\,keV~\cite{Eliseev_2015}) of the decay to the energy of the M1 shell enhances the number of events close to the end point, where the effect of the non-vanishing neutrino mass is more easily measurable. The experimental approach suggested to measure the electrons emitted in the $^{163}$Ho decay was from the start calorimetric,
but yet only in the last decade the technological improvements in low-temperature detectors reached the demanding performances needed for measuring the neutrino mass with sensitivity below $\sim 2$\,eV.

In a calorimetric measurement, the energy released in the decay process is entirely contained into the detector, except for the fraction taken away by the neutrino. This approach eliminates both the issues connected to the use of an external source and the systematic uncertainties arising from decays on excited final states. The most suitable detectors for this type of measurement are low temperature thermal detectors~\cite{Moseley_1984,Fiorini_1983}, where all the energy released into an absorber is converted into a temperature increase that can be measured by a sensitive thermometer directly coupled to the absorber.

The HOLMES experiment~\cite{Alpert_2014} aims to perform a calorimetric measurement of the energy released in the electron capture decay of $^{163}$Ho. The main goal is to reach a neutrino mass statistical sensitivity below 2\,eV proving the potential and the scalability of this technique for a future megapixel experiment. In order to reach this sensitivity HOLMES will collect about $3\cdot 10^{13}$ events by deploying an array of 1024 detectors. The total amount of implanted $^{163}$Ho nuclei will be about $6.5 \cdot 10^{16}$, equivalent to 18\,$\,\mu$g for a total activity of 300\,dec/s/pixel. The target for the instrumental energy and time resolutions are around 3-5 eV FWHM at 2.6\,keV and 20\,$\mu$s respectively. Microcalorimeters based on Transition Edge Sensors (TESs) can easily match these requirements. Considering the large number of pixels and the high event rate, a large multiplexing factor and a large bandwidth are needed. To fulfill these requirements, HOLMES will exploit recent advances on microwave multiplexing ($\mu$mux)~\cite{Noroozian_2013}. 

\section{TES detectors design}
\begin{figure}[!t]
\centering 
\includegraphics[width=0.45\textwidth,clip]{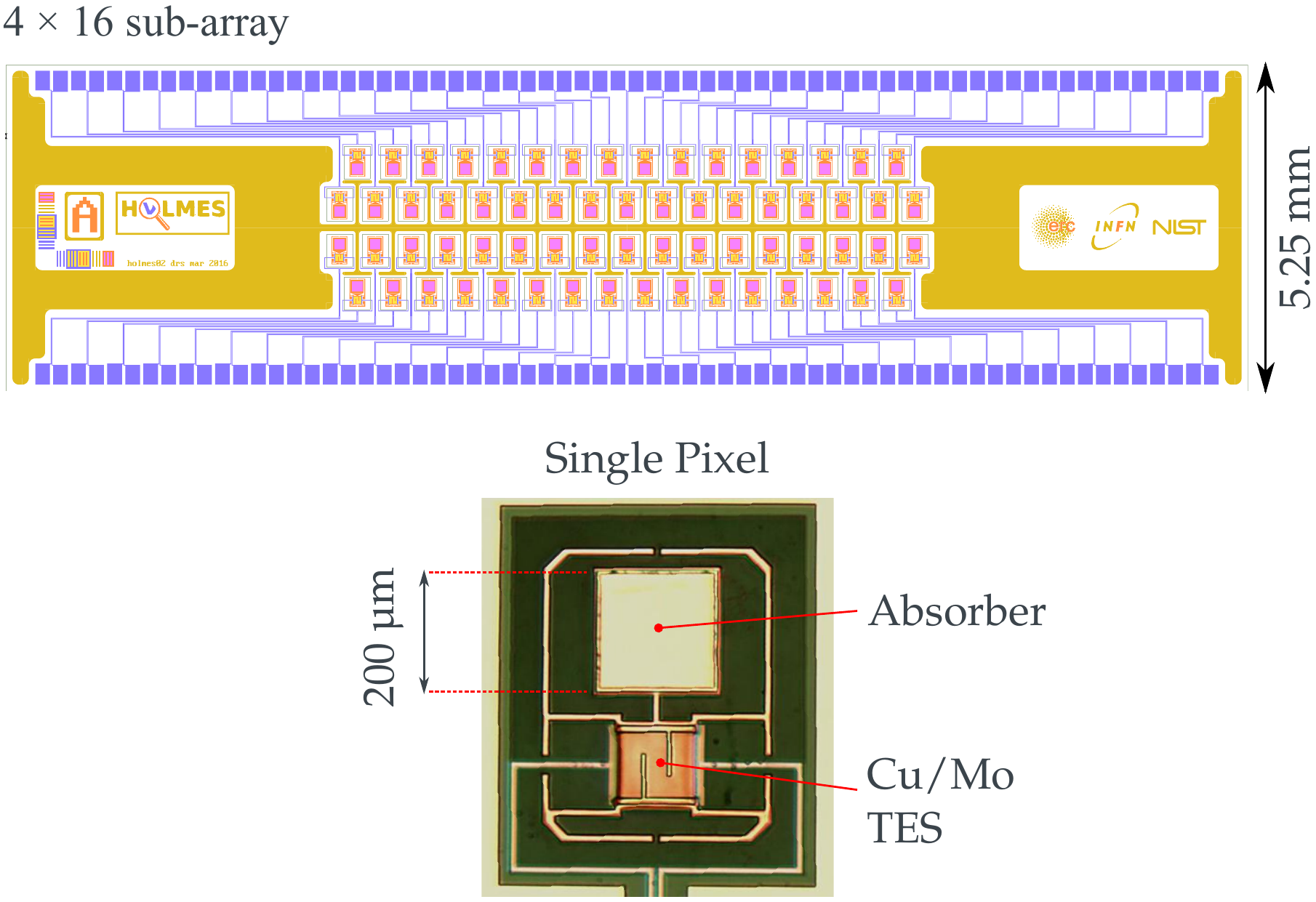}
\caption{\label{fig:stes} (top) The HOLMES chip featuring 63 TESs each; (bottom) Optical microscope image of the HOLMES Mo/Cu TES with Au absorber placed alongside (side-car design).}
\end{figure}

HOLMES uses TES microcalorimeters coupled to gold absorbers designed for soft X-ray spectroscopy with fast response and high energy resolution. A single TES is a $125\times 125\,\mu\mbox{m}^2$ Mo/Cu bilayer, designed to have a transition temperature of $T_c=100$\,mK. A $200\times 200\,\mu\mbox{m}^2$ gold absorber is placed aside the TES to avoid proximity effect between the gold and the sensor itself (\textit{side-car} design, Fig.~\ref{fig:stes}, bottom). The absorber thickness is $2\,\mu\mbox{m}$ in order to provide full containment of the 99.99\% (96.73\%) of the most energetic electrons (photons) produced in the decay~\cite{Nucciotti_2018}. The entire structure is suspended on 500\,nm thick low-stress SiN$_x$ membrane preventing phonons to escape into the silicon substrate. 

The requirement of high pile-up discrimination ability sets strict constraints on the detector response. At first order, the decay time is set by the ratio between the thermal capacity $C$ of the absorber and the thermal conductance $G$ towards the silicon substrate that acts as a constant temperature thermal bath. Since the heat capacity is constrained by the dimensions of the absorber, for the full containment of the released energy, the only way to decrease the pulse decay time is to increase $G$. The thermal conductance $G$ is increased by the addition of a thermal radiating perimeter that increases the conductance in this 2D geometry~\cite{Hays-Wehle_2016}. This thermalizing perimeter increases the thermal conductance $G$ with negligible increase in the total heat capacity. With this design $G$ can be tuned from 40\,pW/K up to 1\,nW/K. The target thermal conductance for HOLMES is around 600\,pW/K, value that guarantees a total pulse time below 200\,$\mu$s and hence a very short \textit{dead time}.

This pixel configuration ensures an energy resolution better than 5\,eV at 5.9\,keV\cite{Puiu_2019}, and an exponential pulse rise time in the range from 10 to 20\,$\mu$s~\cite{Puiu_2018}. With this rise time and with a sampling rate of at least 500 kHz, it is possible to obtain a time resolution better than 3\,$\mu$s exploiting discrimination algorithms based on singular value decomposition~\cite{Alpert_2015} or Wiener filtering~\cite{Ferri_2016}.  

The TES pixels are closely packed in a $4 \times 16$ linear sub-arrays (Fig.~\ref{fig:stes}, top). The design aims to minimize the signal bandwidth limitations due to stray self-inductance of the readout leads; to minimize the signal cross talk due to mutual inductance between readout lines; and to maximize the geometrical filling for an optimal implantation efficiency. These $4 \times 16$ linear sub-arrays are developed in two versions very similar but with minor differences in order to  match a different procedures to release the membrane: Silicon KOH anisotropic wet etching and Silicon Deep Reactive Ion Etching (DRIE). In this latter case, the detectors are placed in a more compact disposition, increasing in this way the geometrical efficiency of the $^{163}$Ho implantation. Each HOLMES sub-array includes 64 TESs and 16 of them will cover all the 1024 foreseen pixels.

\section{TES detectors fabrication}\label{sec:fab}
The TES detectors of HOLMES have to undergo two different fabrication steps~\cite{Orlando_2018}. The detectors are produced at NIST, where the TES, the copper structure and the first part ($1\,\mu$m thickness) of the absorber are evaporated atop of the SiN membrane. All the detectors but the absorbers, are covered in photoresist and shipped to INFN Genova, where the Holmium implantation is made. Holmium is extracted from a metallic target, ionised, accelerated towards the TES array and implanted in the uncovered absorbers~\cite{DeGerone_2019, Gallucci_2020}. This implantation system is composed of an argon penning sputter-based ion source containing $^{163}$Ho, an acceleration section, dipole magnet mass analyzer to select only the isotope of interest, a focusing electrostatic triplet, to refocus the beam, and a magnetic XY scanning stage, to defect the beam in the plane perpendicular to its motion. At the end of the implanter the last section is a vacuum chamber (hereafter Target Chamber) that allows a simultaneous Gold evaporation to control the $^{163}$Ho concentration and to deposit a final Gold layer to protect the $^{163}$Ho from oxidizing.

The ion implantation system is under commissioning phase in a dedicated laboratory at the INFN Unit of Genoa (Italy). Tests are currently performed using a copper sputter target instead of a Ho one, with the goal to study the relationship between the beam current and the ion source parameters. The final sputter target will be a sintered compound composed of Ho (5\%) in a metallic mixture of Ti (36\%), Ni (41\%), and Sn (18\%)~\cite{DeGerone_2019}. This new sputtering target including Ho will be tested as soon as the current tests with Copper target will be finished. 

\begin{figure}[!t]
\centering 
\includegraphics[width=0.45\textwidth,clip]{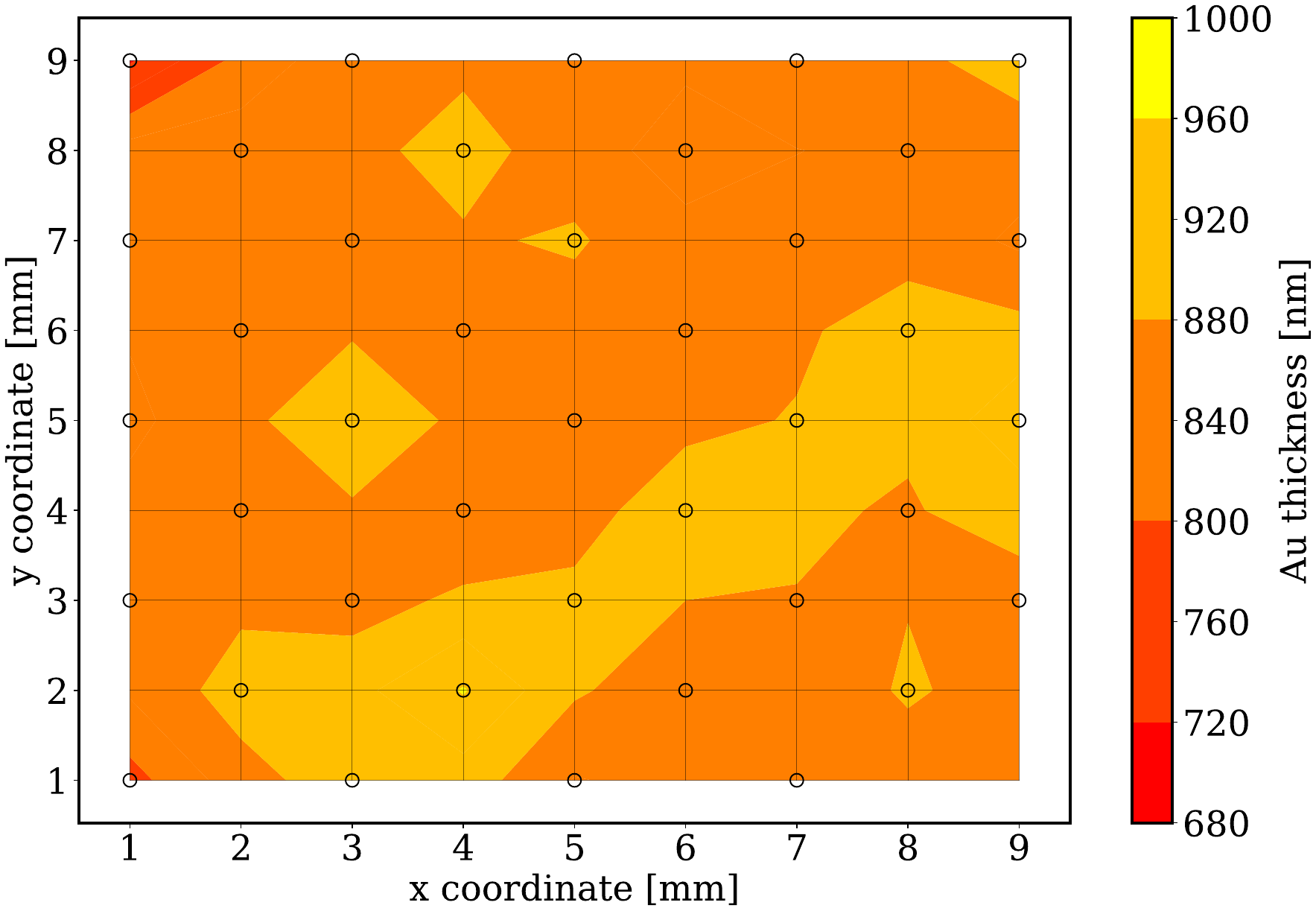}
\caption{\label{fig:sputter} Uniformity of the sputtered Gold inside the developed Target Chamber.}
\end{figure}

The Target Chamber has been developed and tested at Milano Bicocca University. The Target Chamber is equipped with an ion beam assisted sputtering system which will allow a simultaneous Gold deposition to control the $^{163}$Ho concentration and compensate the Gold sputtering caused by the ion implantation, and to deposit the final Gold layer to complete the $^{163}$Ho embedding. The chamber was extensively tested during the past years, evolving from one ion beam microwave source to four ones to increase the deposition rate and improve the uniformity and isotropy of the deposited layer. The uniformity were checked by sputtering Au for $\sim 22$ hours  on a Si slab $1\times1$\,cm$^2$ with a drilled mask with $9\times9$ holes on top. The thickness in the center of the circles were measured with a profilometer finding an average value of $d = 865 \pm 40$\,nm (Fig. \ref{fig:sputter}). Considering a measured deposition rate greater than 50\,nm/hours a Gold thickness of 1\,$\mu$m can be deposited in around 20\,hours. After the deposition the lift-off for the samples is done in hot acetone obtaining a $200 \times 200$\,$\mu$m$^2$ Gold absorber.

The final step of the detector arrays fabrication procedure is the membrane release. At the Milano-Bicocca University the silicon anisotropic wet etching using KOH were successfully developed. This is now the baseline for the two-step fabrication of the first arrays. The design of these sub-arrays is tuned for the potassium hydroxide (KOH) wet etching and this limits the packing of the pixels, because of the sloped side of the openings in the silicon wafer. This design is not optimal for the implantation efficiency which could be improved by about a factor two with a more packed $4\times16$ design. The Deep Reactive Ion Etching (DRIE) process, which provides steep side openings, will allow this highest pixel density. This is currently under study by exploiting an external facility (Trustech, Turin, Italy). After an R\&D phase on dummy samples, tests on real arrays will be performed during 2021. 

Two HOLMES sub-arrays produced with the complete two-step fabrication presented above, though still without the $^{163}$Ho implantation, and with the membrane released by KOH wet etching, were measured during the 2020. The characterization results are presented in the next sections.  

\section{The multiplexed read out}\label{sec:mux}
The readout for the HOLMES detector array is implemented by exploiting the microwave multiplexing scheme~\cite{Giachero_2019}. Each detector is coupled to an rf-SQUID (Superconducting Quantum Interference Devices), which in turn is inductively coupled to a $\lambda/4$ resonator~\cite{Noroozian_2013}. Changes in the TES resistance modulate the current inside the TES circuit. This current variation is then translated into a change in magnetic flux in the rf-SQUID, which in response varies the resonant frequency of coupled resonators. By placing in the same multiplexer chip many microresonators with different resonant frequencies, in the GHz range, and by coupling all the microresonators to a common feedline, it is possible to perform the read out of multiple detectors with a single line. The core of the microwave multiplexing read out is the multiplexer chip. HOLMES uses multiplexer chip ($\mu$mux17) developed at NIST and composed by 33 quarter-wave coplanar waveguide covering 500 MHz in the 4-8 GHz frequency range. Each resonator is designed to have a resonance with bandwidth of 2\,MHz for reading out demodulated signals with a sampling frequency up to 500\,kS/s. The spacing between adjacent resonances is set to 14\,MHz, to assure a negligible cross-talk. 

The microwave read out and multiplexing system developed for HOLMES is based on electronics developed for the readout of microwave kinetic inductance detectors (MKIDs)\cite{Duan_2010,McHugh_2012}. A comb of signals with frequencies tuned to the resonant frequencies of the resonators is digitally generated by a digital processing board, then digital-to-analog converted (DAC) at frequencies of the order of few MHz (base-band, 0-512\,MHz) and finally up-converted in the RF band (4-8\,GHz) by mixing them with a GHz signal (local oscillator—LO). At the output of the resonator chip, these signals are amplified at 4\,K by a low-noise HEMT amplifier, down-converted at room temperature by mixing them with the LO, and then acquired by an analog-to-digital converter (ADC). The resulted digitized waveforms are finally channelized (i.e. individual channel signal recovery) and processed by the digital processing board for real-time reconstruction of the TES signals. This read out chain covers one multiplexer chip with 32 resonators over a 500\,MHz band. One resonator is intentionally left dark to evaluate the SQUID noise performance without the detector contribution.  

A ROACH2 (Reconfigurable  OpenArchitecture Computing Hardware) board~\cite{Hickish_2016} is used to generate the base-band comb signal and to perform the channelization and the digital  processing routines for the real-time demodulation. A semi-custom commercial board, designed to meet HOLMES readout requirements, is used to perform the up-and down-conversion. This board provides a conversion loss of around 7\,dB, value high enough to drive and read out 32 resonators without spoiling their performances. The typical RF bandwidth for the HEMT amplifiers selected for HOLMES is 4\,GHz in the range from 4 to 8 GHz~\cite{LNF_LNC4_8C}. This means that a single HEMT can amplify the signals coming from a series 8 multiplexer chips each one with different frequency bands covering a total of 256 pixel per HEMT. To cover the total 1024 pixels expected for HOLMES, 4 HEMT amplifiers are needed for a total of 32 multiplexer chips, as well as the same number of ROACH2 boards+Up/Down conversion boards, since the readout chain described here covers one multiplexer chip.

The ROACH2 FPGA firmware used to implement the digital signal processing is based on the one developed at NIST for X- and gamma-ray spectroscopy~\cite{Mates_2017,Gard_2018}, with some modifications needed to handle resonators with larger bandwidth. A preliminary 16-channel version exploiting only half of the available DACs/ADCs bandwidth is fully working and it was used in different characterization runs during 2020. The definitive 32-channel version is currently under debugging phase and it will be used for the next measurements. The development of a 64-channel system based on two ROACH2 boards and on two remotely programmable semi-commercial up- and down-conversion circuitry is also currently in progress. This setup will be fundamental for the read out of the first microcalorimeter $4 \times 16$ sub-array with $^{163}$Ho nuclei implanted.

\section{The detector performances}
Two preliminary $6 \times 4$ prototype arrays were entirely produced at NIST with the release of the membrane done with a DRIE process. These array presented slight variations in the pixel perimeter/absorber designs. The goal was to study the different detector responses selecting the best performing solution for the HOLMES requirements, in terms of energy and time resolutions. The results obtained in these tests allowed to identify the optimal detector. The chosen detector achieved an energy resolution of $(4.5 \pm 0.3)$\,eV on the chlorine K$_{\alpha}$ line, at  2.6\,keV, obtained with an exponential rise time and decay of $14\,\mu$s and $200\,\mu$s, respectively~\cite{Alpert_2019}. 

The selected pixel perimeter/absorber were implemented in two $4 \times 16$ sub-arrays that were produced in 2019 following for the first time the entire HOLMES fabrication process presented in Sec. \ref{sec:fab}, with the exception of the implantation of the $^{163}$Ho. One array presents all the pixels identical to the one selected in the previous productions while the second one presents few differences in the sensor/absorber configuration for further tuning of the pixel design. At this stage, the release of the membrane was done with a KOH wet etching. These two arrays have been tested during the 2020 by using the read out system presented in Sec. \ref{sec:mux}. They were installed in a copper holder designed to host 128 channel (2 sub-arrays and 4 multiplexer chips). 8 holders will cover the entire HOLMES in its final configuration (1024 channels). A fluorescence source was employed to test the detectors: this was composed of a primary $^{55}$Fe source faced several targets obtaining their X-ray characteristic emission lines. The total count rate per pixel was around 0.5\,Hz. During the characterization runs performed in 2020 only two multiplexer chips were mounted inside the holder and only half of the pixels were available for acquisition (32+32). Moreover, only 16 pixels per chip were read out at the same time due to the limited bandwidth of the used 16-channel version firmware.

\begin{figure}[!t]
\centering 
\includegraphics[width=0.45\textwidth,clip]{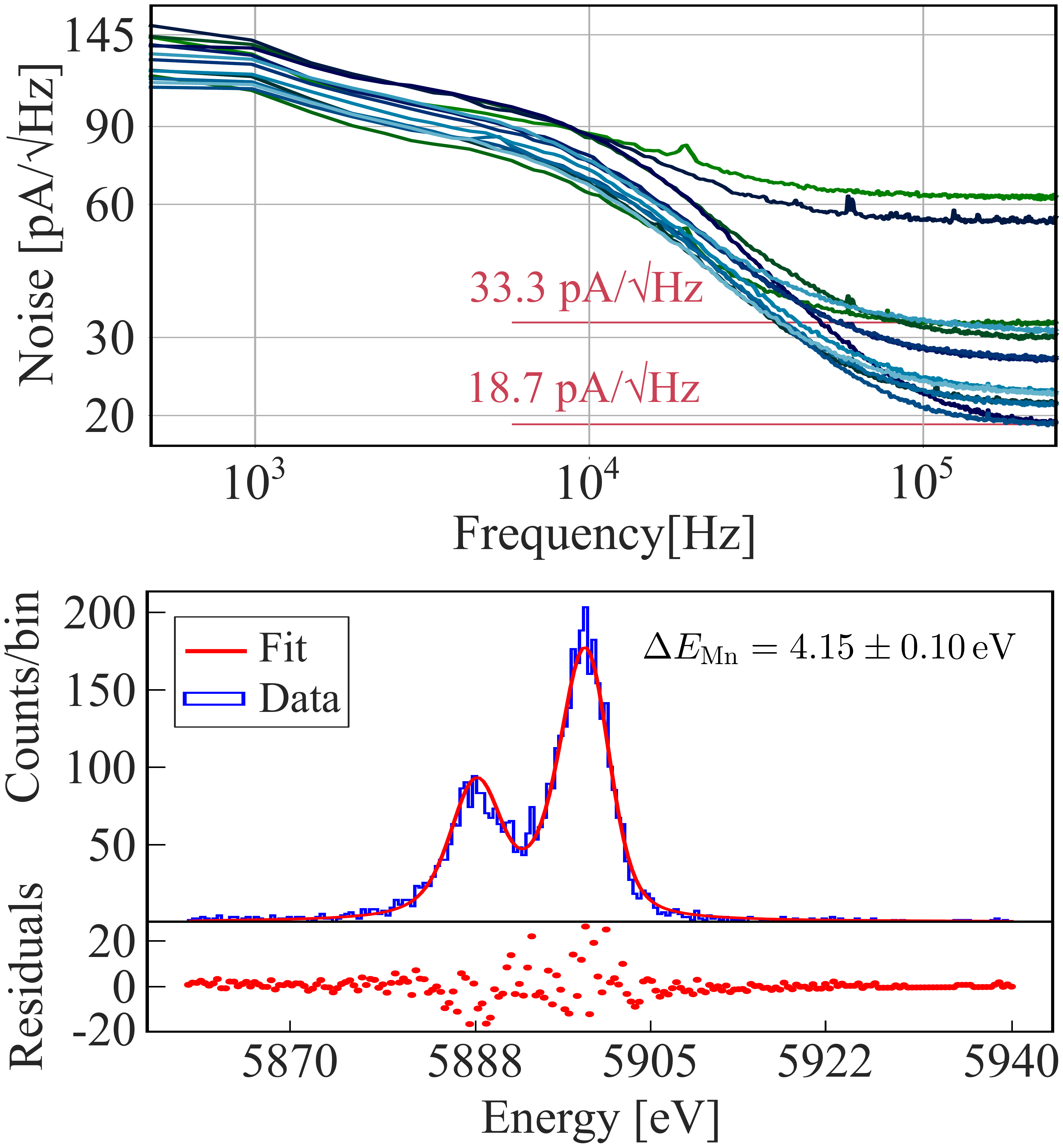}
\caption{\label{fig:plot} (top) Example of noise power spectral density measured for 13 acquired detectors; (bottom) Separation of the K$_{\alpha1}$ and K$_{\alpha2}$ of the Mn, obtained with best performing detector.}
\end{figure}

The obtained results for the array with all the pixels identical are here presented. The read out noise resulted in the $(19-33)\,\mbox{pA}/\sqrt{\mbox{Hz}}$ range (Fig. \ref{fig:plot}, top), compatible with the level obtained in previous measurements~\cite{Alpert_2019, Giachero_2019} and in other applications~\cite{Mates_2017}. Few channel presented an higher noise (around $60\,\mbox{pA}/\sqrt{\mbox{Hz}}$) due to problematic resonators and to no optimal rf-SQUID oscillations. The detectors showed FWMH energy resolutions at 5.9\,keV within the $4 - 6$\,keV range, with a best performing detector having a resolution of $(4.15\pm 0.10)$\,eV (Fig. \ref{fig:plot}, bottom). The measured pulse rise time resulted around $20\,\mu$s, that despite being higher than for previously measured detectors, it still matches the HOLMES requirements. Since the rise time, at the first order, is set by the electrical cutoff of $L /R_0$, where $L$ is the stray inductance and $R_0$ is the resistance of the sensor at the working point, with same $L$ and $R_0$ this larger value may be probably due to the different parasitic impedance in the TES biasing circuit. The tested pixels showed also a longer decay time ($300\,\mu$s vs. $200\,\mu$s) with a large spread. Extrapolating the thermal conductance $G$ through the measurement of the IV curves at different temperatures a mean value of 400\,pW/K was found. This value resulted lower than the expected one of 600\,pW/K, and presented also a dispersion of around $\pm 50$\,pW/K. The cause of this lower $G$ and of its variability is still object of investigation but might be related to the wet etching process. Switching to DRIE might turn out to be the only safe and reliable technique to release the membranes with the grade of reproducibility required by HOLMES. Array processed with DRIE will be tested and characterized during 2021. In parallel, the wet etching procedure will be refined in order to improve the uniformity. Lastly, the pulse amplitude estimated at 5.9\,keV resulted compatible with the ones obtained in the previous productions. Since this amplitude depends on the $\Delta E/C$ ratio, where $C$ is the absorber heat capacity, this means that the developed Target Chamber is properly tuned and the deposited thickness is compatible with the expected one.
  
\section{Conclusion}
Most of the HOLMES detector production processes are being set up, while few of them need some optimizations. To release the membrane either a wet (KOH) or a DRIE etching process can be used to remove the silicon underneath the membrane. The wet process is undergoing the final refinement in order to find the most suitable parameters for the detectors of HOLMES. The target chamber and the gold deposition system is functioning properly and it is currently being set-up and the end of the ion implanter beam. To date, a test of the implanter with a target composed of copper has been performed. Subsequent tests will be performed with targets made of natural holmium and, eventually, of $^{163}$Ho to produce the proper HOLMES detectors. A first measurement with the first $4 \times 16$ sub-array implanted with $^{163}$Ho, and read out with a 64-channel multiplexing system, is expected to start during 2021.

\section*{Acknowledgment}
This work was supported by the European Research Council (FP7/2007-2013) under Grant Agreement HOLMES no. 340321. We also acknowledge the support from INFN through the MARE
project and from the NIST Innovations in Measurement Science program for the TES detector development.

\bibliographystyle{IEEEtran}
\bibliography{asc2020}

\end{document}